\documentclass[12pt]{article}
\usepackage{graphicx}
\input epsf
\usepackage[T2A]{fontenc}
\usepackage[russian,english]{babel}
\begin{document}
\font\bss=cmr12 scaled\magstep 0 \sloppy
\title{The cosmological constant as an eigenvalue \\ of a Sturm-Liouville problem}
\author{Artyom V. Astashenok$^{a,}$\footnote{E-mail: artyom.art@gmail.com},\ Emilio Elizalde$^{b,}$\footnote{E-mail: elizalde@ieec.uab.es, elizalde@math.mit.edu},\ Artyom V. Yurov\,$^{a,}$\footnote{E-mail: artyom\_yurov@mail.ru} \\ \mbox{} \\
\small $^{a}$Theoretical Physics Department, I. Kant Baltic Federal University \\
\small 236041, Nevskogo St., 14, Kaliningrad, Russia \\
\small $^{b}$Consejo Superior de Investigaciones Cient\'{\i}ficas,
ICE/CSIC and IEEC \\
\small Campus UAB, Facultat Ci\`encies, Torre C5-Par-2a pl, 08193 Bellaterra (Barcelona) Spain  }

\date {}
\maketitle

\begin{abstract}

It is observed that one of Einstein-Friedmann's equations has formally the aspect of a Sturm-Liouville problem, and that the cosmological constant, $\Lambda$, plays thereby the role of spectral parameter (what hints to its connection with the Casimir effect). The subsequent formulation of appropriate boundary conditions leads to a set of admissible values for $\Lambda$, considered as eigenvalues of the corresponding linear operator. Simplest boundary conditions are assumed, namely that the eigenfunctions belong to $L^2$ space, with the result that, when all energy conditions are satisfied, they yield a discrete spectrum for $\Lambda>0$ and a continuous one for $\Lambda<0$.
A very interesting situation is seen to occur when the discrete spectrum contains only one point: then, there is the possibility to obtain appropriate cosmological conditions without invoking the anthropic principle. This possibility is shown to be realized in cyclic cosmological models, provided the potential of the matter field is similar to the potential of the scalar field. The dynamics of the universe in this case contains a sudden future singularity.

\end{abstract}

\sloppy

\section{Introduction}

In Einsteinian Gravity, when quantum corrections are taken into account the determination of the
cosmological constant (cc), $\Lambda$, is very much related with vacuum fluctuations and, eventually, the Casimir effect \cite{Book}, in one way or other \cite{eecc}. The problem of the cc is one of the most important and intriguing in modern cosmology and QFT. The traditional way to understand why $\Lambda$ is so small ($\Lambda <10^{-120}$ in Planck units) is ordinarily based on the anthropic principle \cite{anthropic-begin}. Up to now one has no reasonable alternative way to explain this fact in Einsteinian Gravity, unless one moves on to extended models with dynamical dark energy or modified gravities \cite{Odin-Noj-rep-2011}.

Recently Barrow and Shaw \cite{barrow-new-solution} suggested a non-anthropic solution to the problem: by making $\Lambda$ into a field and restricting the variations of the action with respect to it by causality, they managed to obtain an additional Einstein constraint equation. One can say that this approach is based on a different {\em interpretation} of the cosmological constant, notably to consider the cosmological constant to be a field variable.

In the present work we use a somehow similar procedure to calculate the value of $\Lambda$, namely to interpret it as an eigenvalue of a Sturm-Liouville problem, rather than as an integration constant. Our approach is in fact more conservative than the method in \cite{barrow-new-solution} since we use the standard Einstein-Friedmann equations {\em without} any additional constraint equation. Instead, we consider one of Friedmann's equations as a spectral problem and look for a class of boundary conditions which may allow us to actually calculate the corresponding eigenvalues (in this way, our procedure is closely related to standard investigations of the Casimir effect \cite{Bordag-2M-Casimir,Book,eecc} by spectral methods). A most simple condition is to impose that eigenfunctions be elements of $L^2$. As we will show, this choice results in the following interesting consequence: if the universe is filled up with a matter field, $\phi$ (except for $\Lambda$), such that all the strong energy conditions are satisfied, then one gets a point spectrum for positive values of the cosmological constant, and a continuous spectrum for negative ones.

The next step will be to consider  models  which actually result in a positive discrete spectrum with just one {\em single} point. This idea is inspired in the conceptual approach developed in the paper by Linde and Vanchurin \cite{Linde-Vanchurin-1011}: one can  fix all the parameters in the landscape, including the value of the cosmological constant, without using any anthropic considerations, what ``will give us a chance to return to Einstein's dream of a final theory, which may allow us to make sharp and unambiguous predictions despite the abundance of choices''. Clearly, this case of a one-point spectrum is a good  example of a possible  realization of this cherished dream.

An additional bonus of our approach is the surprising connection that will be uncovered between the Steinhardt-Turok cyclic models \cite{cyclic} and Barrow's sudden singularity \cite{sudden}---or singularity of type II according to the classification in \cite{ClassOfSing-Odin}. In short, on one hand, models with Steinhardt-Turok-like cyclic potentials may indeed have much to do with those singularities, and on the other hand, such models exhibit a single-point positive spectrum for $\Lambda$ in the framework of our approach. This striking connection will be a crucial issue in this article.

\section{Friedmann's equation as a Sturm-Liouville problem}

A method for constructing and analyzing exact cosmological
solutions of the Einstein equations based on representing them as
a second-order linear equation (we call this the linearization
method, in what follows) was discussed in \cite{Chervon} (other
methods for constructing exact solutions in cosmology can be found
in \cite{Barrow}-\cite{Barrow2}). Indeed, it is easy to see that,
in the case of the flat Friedmann metric, the third power
of the scale factor $\psi =a^3$ satisfies the equation
\begin{equation}
\frac{d^2\psi}{dt^2}=\frac{9}{2}\left(\rho-p\right)\psi,
\label{a3}
\end{equation}
where $\rho$ is the density and $p$ the pressure of the matter
filling the universe. In this paper we use the system of
units with $8\pi G/3=c=1$. In the case when a minimally coupled
scalar field, $\phi$, with the self-interaction potential $V(\phi)$,
is dominant, and in the presence of a cosmological constant with
 density $\Lambda$, Eq.~(\ref{a3}) formally coincides with the
Schr\"{o}dinger equation
\begin{equation}
\frac{d^2\psi}{dt^2}=(U-\lambda)\psi, \label{Schr}
\end{equation}
where the potential is $U(t)=9V$, and the spectral parameter is
$\lambda=-9\Lambda$. In (\ref{Schr}), the quantity $V$ is assumed
to be a function of time: $V(t)=V(\phi(t))$, which has been called {\it the
history} of the potential in \cite{Chervon}. Giving an explicit
form $U(t)$ together with the corresponding boundary conditions
allows to find the general solution of Eq.~(\ref{Schr}). An important
consequence of this investigation is that the regime is
independent of, or just weakly dependent on the type of the potential,
what is quite significant for the whole theory. Unfortunately,
the ubiquitous problem of the end of inflation turns out to be substantially
more difficult here, and solving it in the framework of the approach
described above apparently involves additional assumptions. (The
authors of \cite{Chervon} proposed modifying the potentials to
make them depend on the temperature. Stablishing a
Friedmann regime can then be described as a phase transition in
the matter state of the early universe). The study of Eq.~(\ref{Schr})
in its applications to cosmology was continued in \cite{Yurov1} and
\cite{Yurov2}, where the Darboux transformation was used to
construct new exact solutions (a similar technique was used in
\cite{Yurov3} to obtain exact solutions on the brane and on the
encompassing space carrying an orbifold structure).

We consider the Einstein equations in the Friedmann metric,
\begin{eqnarray}
  \frac{\dot{a}^2}{a^{2}} &=& \rho-\frac{k}{a^{2}},
  \label{Ein1}\\
  \frac{\ddot{a}}{a} &=& -\frac{1}{2}\left(\rho+3p\right) \label{Ein2},
\end{eqnarray}
and assume that $a=a(t)$, $p=p(t)$, $\rho=\rho(t)$ is a solution of
these equations, for $k=0$. Then, the function $\psi_{n}=a^{n}$ is a
solution of the Schr\"{o}dinger equation
\begin{equation}
{\ddot \psi_{n}}=U_{n}(t)\psi_{n}, \label{2}
\end{equation}
where the ``potential'' is
\begin{equation}
U_{n}(t)=n^{2}\rho-\frac{3n}{2}(\rho+p). \label{Un}
\end{equation}
If the universe is filled up with a scalar field $\phi$ with the
Lagrangian $L=\frac{\dot{\phi}^{2}}{2}-V(\phi)$, then
\begin{equation}
U_{n}=\frac{n(n-3)}{2}\dot{\phi}^{2}+n^{2}V(\phi). \label{Un-1}
\end{equation}

The effectiveness of the method presented in
\cite{Chervon} (and developed in \cite{Yurov1} and
 \cite{Yurov2}), precisely consists
in reducing a complex nonlinear problem into a linear equation. This
allows to obtain a complete set of two-parameter solutions, which exhibit
inflationary behavior under very general assumptions. The fact
that $U$ basically coincides with $V$ was not used anywhere in
those papers and consequently played no role there. In a similar way, we here
consider a generalization of this method to arbitrary $n$.

Further to this point, the physical meaning of the ``potential'' $U_3$ is
clear only for a universe filled with a scalar field. Indeed, if we
consider a universe in which, for instance, electromagnetic
radiation is dominant, the physical meaning of the quantity $U_3$
becomes again ambiguous.

We note that if a solution of (\ref{Schr}) is known, then we can
use Eqs.~(\ref{Ein1}) and (\ref{Schr}) to find the scalar field
\begin{equation}
\phi(t)=\pm\frac{\sqrt{2}}{\sqrt{3n}}\int
dt\sqrt{\frac{\dot{\psi}_{n}^{2}}{\psi_{n}^{2}}-U_{n}+\lambda_{n}},\label{phit}
\end{equation}
 and the potential
\begin{equation}
V(t)=\frac{1}{3}\left[\frac{U_{n}}{n}+\frac{3-n}{n^{2}}\left(\frac{\dot{\psi}_{n}^{2}}{\psi_{n}^{2}}+\lambda_{n}\right)
\right]. \label{Vt}
\end{equation}
In principle we can obtain the dependence $V=V(\phi)$ from these expressions,
although it is clearly not always possible to do this explicitly. For example, if $n=1$ and
$$
U(t)=4t^2\left(4t^4-3\right),
$$
then
$$
\rho=16t^6,\qquad p=8t^2(1-2t^4), \qquad
w=-1+\frac{1}{2t^4},\qquad H=-4t^3,
$$
and
$$
V(\Phi)=4\sqrt{2}\Phi^3+4\sqrt{3}\Phi^2,
$$
after performing the Bogoliubov transformation $\phi=1/\sqrt{6}+\Phi$.

In the general case, a solution of (\ref{2}) has the form
\begin{equation}
\Psi_{n}=c_{1}\psi_{n}+c_{2}\hat{\psi}_{n},\label{general}
\end{equation}
where $\hat{\psi}_{n}$ is a linearly independent solution with the
same potential:
\begin{equation}
{\hat\psi}_n(t)=\psi_n(t)\int^t \frac{dt'}{\psi_n^2(t')}\equiv
\psi_n(t)\xi(t). \label{xi}
\end{equation}
Equation (\ref{general}) allows to prove the following statement.
\medskip

\noindent {\bf Assertion}. Let $a=a(t)$ be a solution of (\ref{Ein1}) and
(\ref{Ein2}) for $k=0$ and the corresponding $\rho$ and $p$. Then,
the three-parameter function $a_n=a(t;c_1,c_2,n)$, of the form
\begin{equation}
a_n=a\left(c_1+c_2\int\frac{dt}{a^{2n}}\right)^{1/n}, \label{an}
\end{equation}
is a solution of Eqs.~(\ref{Ein1}) and (\ref{Ein2}) for the new energy
density $\rho_n$ and pressure $p_n$ satisfying the condition
\begin{equation}
n^2\rho_n-\frac{3n}{2}\left(\rho_{n}+p_{n}\right)=
n^2\rho-\frac{3n}{2}\left(\rho+p\right). \label{inv}
\end{equation}

Finally, if we assume that the universe contains a nonzero
vacuum energy with density $\rho_{_\Lambda}c^2$, in addition
to the matter fields, then Eq.~(\ref{2}) takes the form of the
spectral problem
\begin{equation}\label{Scr-11}
 \ddot{\psi}_{n}=(U_{n}(t)-\lambda_{n})\psi_{n},
\end{equation}
where the spectral parameter is
$\lambda_{n}=-n^{2}\rho_{_\Lambda}$. Just as for Eq.~(\ref{Schr}), we can consider a problem for the eigenvalues and
eigenfunctions of Eq.~(\ref{Scr-11}), if we specify homogeneous
initial conditions. As noted in \cite{Chervon}, Eq.~(\ref{Schr})
has the form of a quantum mechanical problem with a discrete
spectrum. The fact that each solution of this kind only admits a bounded
or countable set of allowed values of the cosmological constant
(if we specify homogeneous initial conditions) may help a lot to clarify the
question of the actual value of the cosmological constant. It is clear that, in the case $n=1$, the SEC results in the condition $U(t)<0$. If one would like to obtain the discrete spectrum of $\Lambda$ for such potentials, one would need to suggest $U(t)\to 0$ at $|t|\to\infty$ and this readily means that one can get the discrete spectrum for positive values of $\Lambda$ only.

This leads one to ask the following question: Is it possible to obtain models with a discrete spectrum containing just one single point?

For answer one need to use some boundary condition on $\psi_{n}$. In this article we suggested that $\psi_n$ is element of square-integrable functional space $L_2$. This is just the hypotheses about boundary condition. Therefore, our aim looks as follows: to investigate the class of cosmologies which results in just one allowed value of vacuum energy, on conditions that $\psi_n$ is element of $L_2$. We'd like to understand rather physical ground which may results in this picture.
In the issue, we have obtained two results, two cosmologies which admits the possibility under consideration:

1) The universe with some new form of sudden future singularity (Sec. 3).

2) The universe without exact sudden future singularity (rather with ``smoothed singularity'') but filled with scalar field $\phi$ with deep narrow self-acting potential $V(\phi)$ (Sec. 4). Such potential  is  very similar with cyclic potential of Steinhardt-Turok (ST) except that the ST cyclic potential $V(\phi)$  has a different asymptotic behavior at  $\phi\to \pm\infty$. Therefore, to obtain ST potential (and breathe the life into our formal model!) one need  to modified the model from the Sec. 4. We do it in the Sec. 5.

Thus, we have two cosmologies which allows one to obtain the extreme form of non-anthropic solution of the problem of cosmological constant. Remarkably, these cosmologies are not absolutely new, speculative models; vice versa, these cosmologies are connected with two well known and popular (at present) models; Barrow sudden future singularities and Steinhardt-Turok cyclic models.
And last, but not least: these two models are connected to each other by the following way - the sudden model (Sec. 3) may be obtained from the ``ST''  models by the  limiting process.

\section{New class of sudden future singularities}

We assume that the matter field contains baryons, that there is also radiation and dark matter, and
that the dark matter leads to a sudden future singularity at $t=t_{s}$, namely
\begin{equation}\label{sfs}
p_{s}(t_{s})=p_{s}=+\infty, \quad
\rho_{s}(t_{s})=\rho_{s}<\infty,\quad 0<a(t_{s})<\infty.
\end{equation}
For the realization (\ref{sfs}), let us choose the dark matter pressure to be of the form:
\begin{equation}\label{p-sfs}
p_{DM}(t)=\widetilde{p}_{DM}(t)+\frac{2}{3}\alpha^{2}\delta(t_{s}-t).
\end{equation}
Hence, $\widetilde{p}_{DM}(t)$ is the ``regular'' part  of the function
$p_{DM}(t)$, while $\alpha^{2}$ is a positive constant. From the equation of state for the dark matter, $p_{DM}=w_{DM}\rho_{DM}$, in which
$w_{DM}=0, t\neq t_{s}$, it follows that
$\widetilde{p}_{DM}(t)=0$. Then, for the energy density and pressure, we have
\begin{equation} \label{rho-sfs}
\rho=\lambda^{2}+\rho_{DM}, \quad
p=-\lambda^{2}+\frac{2}{3}\alpha^{2}\delta(t_{s}-t).
\end{equation}
The contribution of the dark matter decreases with time for $t<t_{s}$. For $t=t_{0}$,
$\rho_{DM}/\lambda^{2}<0.5$ and we can neglect $\rho_{DM}$ in the equations. Therefore,
Eq.~ (\ref{Scr-11}) can be written in the form
\begin{equation}\label{8}
\ddot{\psi}=(\lambda^{2}-\alpha^{2}\delta(t_{s}-t))\psi,
\end{equation}
Let us now assume that the function $\psi$ is continuous at the point $t=t_{s}$. The first derivative of the scale factor has a gap at the time of occurrence of the sudden future singularity, and
\begin{equation}\label{jump-1}
\dot{a}_{s}^{(+)}-\dot{a}_{s}^{(-)}=-\alpha^{2}a_{s},
\end{equation}
where $\dot{a}_{s}^{(\pm)}=\lim \limits_{t\to t_{s}\pm 0}
\dot{a}(t)$ and $a_{s}=a(t_{s})$. From the physical point of view, this gap corresponds to a density jump in the singularity. One gets,
\begin{equation}\label{jump}
\delta\rho=\frac{\dot{a}_{s}^{(+)2}-\dot{a}_{s}^{(-)2}}{a_{s}^{2}}=
-\frac{\alpha^{2}(\dot{a}_{s}^{(+)}+\dot{a}_{s}^{(-)})}{a_{s}}.
\end{equation}
For simplicity one can set  $\delta\rho=0$. Therefore, if the universe expands before the time $t=t_{s}$, then at time $t=t_{s}$ contraction begins. It is clear that the solutions for $t<t_{s}$ and $t>t_{s}$ cannot be connected. But it is interesting to investigate this model further.

The condition (\ref{jump-1}) at $\delta\rho=0$ leads to
$$
H_{s}^{2}=\displaystyle{\left(\frac{\dot{a}_{s}^{(\pm)}}{a_{s}}\right)^{2}}=\frac{\alpha^{4}}{4}.
$$
Therefore, the density at the time of SFS is given by
$\rho_{s}=\alpha^{4}/4+k/a_{s}^2$. One notes that for the open universe the following condition should be satisfied
\begin{equation}\label{condopen}
\alpha^2-2a_{s}^{-1}>0.
\end{equation}
Solving (\ref{8}) with the conditions $\delta a_{s}=0$ and $\delta\rho_{s}=0$ yields
\begin{equation}
a_{-}(t)=a_{s}\exp{\frac{\alpha^{2}}{2}(t-t_{s})}, \quad
a_{+}(t)=a_{s}\exp{\frac{\alpha^{2}}{2}(t_{s}-t)}.
\end{equation}
Therefore, the value of $\Lambda$ is unique and, for the vacuum energy density, we have
\begin{equation}
\Lambda=\frac{\alpha^4}{4}, \label{LLL}.
\end{equation}
Condition (\ref{condopen}) can be written as follows
$$
a_{s}^2>\frac{1}{\Lambda}.
$$

\section{Smoothed SFS}

Strictly speaking, the model in the previous Section has no physical sense. This is because one had to consider two different universes (with positive and negative Hubble roots, respectively) separated by the singularity. One needs to involve a smoothed ``singularity'', such that $|p|\to |p_s|\gg 1$, but $|p_s|<\infty$ and $\rho_s<\infty$.
Let us take
\begin{equation}\label{SSFS}
U_{\kappa}(t)=-\frac{\alpha^{2} \kappa}{2\cosh^{2}(\kappa t)}.
\end{equation}
If $\kappa\rightarrow\infty$, we have
\begin{equation}
U_{\kappa}(t)=\left\{\begin{array}{ll}
0,\quad t\neq 0,\\
-\infty, \quad t=0.\end{array}\right. ,
\end{equation}
and taking into account that
$$
\int_{-\infty}^{+\infty}U_{\kappa}(t)dt=-\alpha^{2},
$$
one concludes that
$$
\lim_{\kappa\rightarrow\infty} U_{\kappa}(t)=-\alpha^{2}\delta(t).
$$
The potential as a function of time, for various values of $\kappa$, is depicted on Fig. 1. Therefore, for $\kappa=\infty$ the potential (\ref{SSFS}) leads to a sudden future singularity, as considered above. We assume that $\kappa>>0$ but $\kappa<+\infty$. The solution of (\ref{Scr-11}), for the potential (\ref{SSFS}), can be written in the form
\begin{equation}\label{psi-SSFS}
\psi(t)=(1-\xi^{2})^{\epsilon/2}F\left(\epsilon-s, \epsilon+s+1,\epsilon+1,\frac{1-\xi}{2}\right).
\end{equation}
Here $F$ is the hypergeometric function, and we have used the following definitions:
$$
\xi=\tanh(\kappa t),\quad \epsilon=\frac{\lambda}{\kappa},\quad s=\frac{1}{2}\left(-1+\sqrt{1+\frac{2\alpha^{2}}{\kappa}}\right).
$$
The condition $\kappa\rightarrow\pm\infty$ corresponds to $\xi\rightarrow \pm1$. Finiteness of the function requires that
$$
\epsilon-s=-N,\quad N=0,1, \ldots
$$
and, therefore, for the spectrum
\begin{equation}\label{spectrum}
\lambda_{n}=\frac{\kappa}{2}\left(\sqrt{1+\frac{2\alpha^{2}}{\kappa}}-1\right)-\kappa N.
\end{equation}
For $\kappa\rightarrow\infty$, one can easily show that
$$
\lambda_{N}=\alpha^{2}/2-\kappa N\rightarrow-\infty, \quad \mbox{if} \quad N\neq0.
$$
But there is the condition that $\epsilon>0$ (as for $\epsilon<0$ the function (\ref{psi-SSFS}) diverges at $\xi\rightarrow\pm1$). Therefore, $\lambda>0$ and from Eq.~(\ref{spectrum}) one has
\begin{equation}\label{condition}
\left(\sqrt{1+\frac{2\alpha^{2}}{\kappa}}-1\right)>2N.
\end{equation}
For sufficiently large $k$ this inequality can be satisfied for $N=0$ only. The remaining single eigenvalue is given by
$$
\lambda_{*}=\frac{\kappa}{2}\left(\sqrt{1+\frac{2\alpha^{2}}{\kappa}}-1\right), \quad \mbox{for}\quad \kappa>\alpha^{2}/4,
$$
and, in this limit, one gets for the cosmological constant this same value.

For the eigenfunctions (\ref{psi-SSFS})
\begin{equation}
\psi(t)=(1-\xi^{2})^{\epsilon/2}F\left(0, 2\epsilon+1,\epsilon+1,\frac{1-\xi}{2}\right)\equiv (1-\xi^{2})^{\epsilon/2}=\cosh(\kappa t)^{-\epsilon},\quad \epsilon=\frac{1}{2}\left(\sqrt{1+\frac{2\alpha^{2}}{\kappa}}-1\right),
\end{equation}
and the Hubble parameter is
\begin{equation}
H=\frac{d\ln\psi(t)}{dt}=\lambda(\kappa)\tanh(\kappa t),\quad \lambda(\kappa)=\frac{\kappa}{2}\left(\sqrt{1+\frac{2\alpha^{2}}{\kappa}}-1\right).
\end{equation}
For the matter pressure and density, we get
$$
\rho_{m}=-\frac{\lambda(\kappa)^{2}}{\cosh^{2}(t\kappa)},\quad p_{m}=\frac{\alpha^{2} \kappa+\lambda^{2}(\kappa)}{3\cosh^{2}(\kappa t)}.
$$
The evolution of the scalar field can be obtained from the condition $\rho_{m}+p_{m}=\dot{\phi}^{2}$,
\begin{equation}\label{phiSSFS}
\phi=\phi_{0}\pm\frac{2}{\kappa}\sqrt{\frac{\alpha^{2} \kappa-2\lambda^{2}(\kappa)}{3}}\arctan(\exp(\kappa t)).
\end{equation}
For simplicity, we choose the ``+'' sign in (\ref{phiSSFS}) and set $\phi_{0}=0$. The potential of the scalar field is similar to the sine-Gordon potential:
\begin{equation}
V(\phi)=-\frac{4\lambda^{2}(\kappa)+\alpha^{2} \kappa}{3}\sin^{2}(2\Phi), \quad \Phi=\frac{\kappa}{2}\sqrt{\frac{3}{\alpha^{2} \kappa-2\lambda^{2}(\kappa)}}\phi
\end{equation}
The potential of scalar field as function of $\phi$ is depicted on Fig. 2. For large $k$ the scalar field changes sufficiently slowly. The corresponding potential at $t=0$, for $k>>1$, has a deep well:
$$
V(t=0)\approx-\alpha^{2} \kappa/3.
$$
\begin{figure}
\begin{center}
\includegraphics{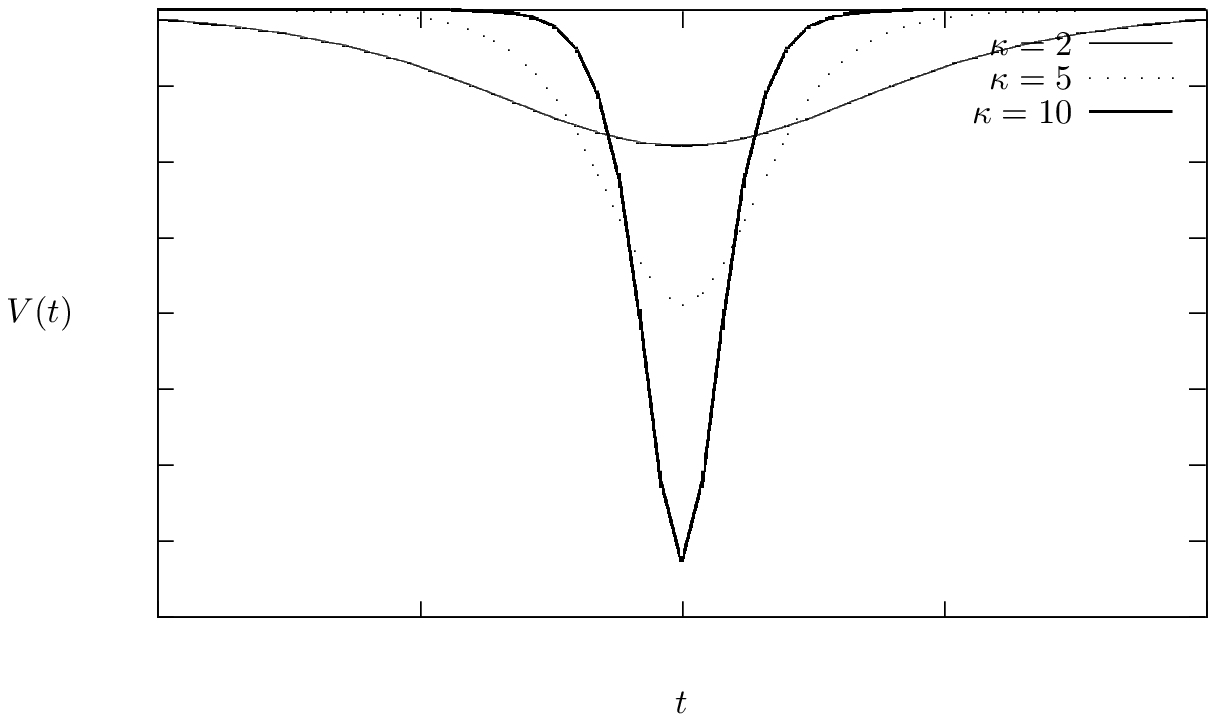}
\caption{The scalar field potential as a function of time for various values of $\kappa$. The depth of the well grows, while its width decreases, with $\kappa$.}
\end{center}
\end{figure}
\begin{figure}
\begin{center}
\includegraphics{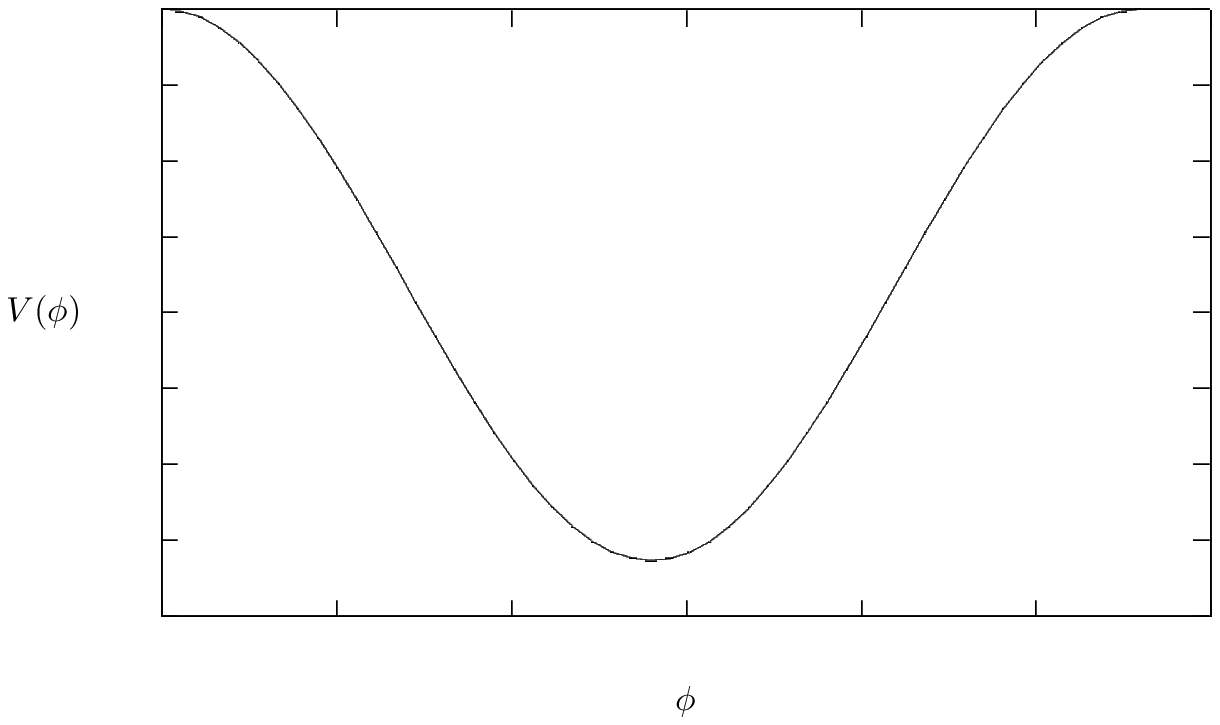}
\caption{The scalar field potential as a function of $\phi$. The scalar field rolls up from 0 to $\phi_{max}=3^{-1/2}\pi\kappa^{-1}(\alpha^{2}\kappa-2\lambda^{2}(\kappa))^{1/2}$ }
\end{center}
\end{figure}

\section{Asymmetric potential}

The potential considered in previous section is very similar with cyclic potential except that the Steinhardt-Turok cyclic potential $V(\phi)$  has a different asymptotic behavior at  $\phi\to \pm\infty$. One can easily construct this potential starting from an appropriate $U(t)$.
Let us here investigate the case when the potential $U(t)$ is taken with the form of a potential well
$$
U(t)=0, \quad t<-\tau,
$$
$$
U(t)=-U_{0}, \quad -\tau \leq t \leq 0,
$$
$$
U(t)=u_{0}, \quad t>0.
$$
Here $U_{0}, u_{0}$, and $\tau$ are positive constants. One can perform the parametrization $U_{0}=-\lambda\cosh^{2}\xi$,
$u_{0}=-\lambda\sinh^{2}\eta$, where $\xi$ and $\eta$ are new constants, and write the solution (\ref{Scr-11}) in the form

1) $t<-\tau$: \begin{equation}\label{area1}
\psi_{1}=C_{1}\exp(\sqrt{-\lambda} t);
\end{equation}

2)$-\tau\leq t \leq 0$:
\begin{equation} \label{area2}\psi_{2}=C_{2}\sin(\sqrt{-\lambda}
\sinh\xi \cdot t+\delta);
\end{equation}

3) $t>0$:
\begin{equation} \label{area3} \psi_{3}=C_{3}\exp(-\sqrt{-\lambda}
\cosh\eta \cdot t).
\end{equation}
Here $C_{i}$ and $\delta$ are constants. We consider the solutions which are finite at $t\rightarrow\pm\infty$.

The function $\psi$ and its first derivative should be continuous at the points $t=\tau$ and $t=0$. These conditions can be written as
$$
\frac{\dot{\psi_{1}}(-\tau)}{\psi_{1}(-\tau)}=\frac{\dot{\psi_{2}}(-\tau)}{\psi_{2}(-\tau)},\qquad
\frac{\dot{\psi_{2}}(0)}{\psi_{2}(0)}=\frac{\dot{\psi_{3}}(0)}{\psi_{3}(0)}.
$$
The first condition yields $\tan(\sqrt{-\lambda}\sinh\xi \cdot \tau-\delta)=-\sinh\xi$,
and the second  $\tan\delta=-\frac{\sinh\xi}{\cosh\eta}$. Solving these equations for $\sqrt{-\lambda}$ and $\delta$, one gets
\begin{equation} \label{cond}
\tan(\sqrt{-\lambda}\sinh\xi\cdot\tau)=\frac{\sinh\xi(1+\cosh\eta)}{\sinh^{2}\xi-\cosh\eta}
\end{equation}
and
\begin{equation}\label{cond-2}
\tan\delta=-\frac{\sinh\xi}{\cosh\eta}.
\end{equation}
We now take the limits $\xi\rightarrow\infty$, $\tau\rightarrow0$, and assume that
$$
U_{0}\tau=-\lambda\cosh\xi\cdot\tau=\sigma,
$$
where $\sigma=\mbox{const}$. This limit corresponds to increasing the depth of the potential well at $-\tau<t<0$ and narrowing it so that its area remains constant. We also assume, for simplicity, that $\eta<<\xi$.

Then, the condition (\ref{cond}) can be written as
$$
\tan z=\frac{1+\cosh\eta}{\sigma}\sqrt{-\lambda} z,\quad z=2\sigma\exp(-\xi)/\sqrt{-\lambda}.
$$
From the condition $z<<1$, that is $\tan
z\approx z$, the corresponding one on the value of $\sqrt{-\lambda}$ follows:
\begin{equation}\label{lambda}
    \sqrt{-\lambda}=\frac{\sigma}{1+\cosh\eta}.
\end{equation}
Thus, the value of the cosmological constant is unique and it is completely determined by the width and the depth of the well, and by the parameter $\eta$, too.

Using Eqs.~(\ref{phit}), (\ref{Vt}) we can obtain the evolution of the scalar field $\phi(t)$ and potential $V(\phi)$. For simplicity we assume that $k=0$. For $n=3$, from these equations we have, in the case of flat spacetime

(i) $t<-\tau$:
\begin{equation}
\phi_{1}=\phi_{10}, \quad V(\phi)=0
\end{equation}

(ii) $-\tau<t<0$:
\begin{equation} \label{solution2}
\phi_{2}=\phi_{20}+\sqrt{\frac{2}{3}}\ln|\tan(\sqrt{-\lambda}\sinh\xi\cdot
t/2+\delta/2)|, \quad V(\phi)=\frac{-\lambda\sinh^{2}\xi}{3}\left(\cosh
(\sqrt{6}(\phi_{2}-\phi_{20}))-2\right)+\lambda,
\end{equation}

(iii) $t>0$:
\begin{equation}
\phi_{3}=\phi_{30},\qquad V(\phi)=-\lambda\sinh^{2}\eta.
\end{equation}

In fact we have here used an unphysical sharp-cornered potential well. In a more realistic situation one needs use a potential well with ``rounded edges''. In this case, the potential of the scalar field $V(\phi)$ becomes a smooth function of $\phi$. Now, let us consider Eqs.~(\ref{solution2}) and put that $U_{0}>>0$ then $\delta=Pi/2$. Since, at $t=-\tau$, $\phi=\phi_{20}-\sqrt{2U_{0}/3}\tau/4$ and, at $t=0$, $\phi\approx\phi_{20}$, one must conclude that
$$
V(\phi_{20}-\delta\phi)=-\frac{U_{0}}{3}(1-U_{0}\tau^{2}),\quad V(\phi_{20})=-\frac{U_{0}}{3}.
$$
Therefore, in the interval $-\tau<t<0$ the potential exhibits a very narrow and deep peak, of width $\delta\phi=\sqrt{2U_{0}/3}\tau/4$ and depth $-U_{0}/3$. It is our conclusion that in a more realistic case one will obtain a potential for the scalar field similar to the potentials in cyclic models.

It is interesting to consider the conditions (\ref{cond}) and (\ref{cond-2}) for a well with finite depth and width. The parameters $\xi$ and $\eta$ can be expressed through the spectral parameter, the depth of the well and the height of the barrier. The width of the well, $\tau$, can be written as
$\tau=\beta/\sqrt{U_{0}}$, where $\beta$ is a new parameter. Then, Eq.~(\ref{cond}) turns into
\begin{equation}\label{trancend}
\tan(\beta\sqrt{1-1/x})=\frac{\sqrt{x-1}(1+\sqrt{1+u_{0}x/U_{0}})}{x-1-\sqrt{u_{0}x/U_{0}+1}},
\end{equation}
where $x=-U_{0}/\lambda$. For $0<\beta<\pi/4$, Eq.~(\ref{trancend}) has only one solution. The root exists provided the condition $\tan\beta>\sqrt{u_{0}/U_{0}}$ is fulfilled.

For illustration, choose $\beta=0.5$,
$U_{0}=1$, and $u_{0}=0.2$, in which case $\tau=0.5$. The numerical solution (\ref{trancend}) yields $x=191.393$ and, therefore, $-\lambda=0.00522$. The parameters $\xi$ and $\eta$
are equal to 3.319 and 2.522, respectively. From Eq.~(\ref{cond}) it follows that $\delta=-1.146$.

\section{Conclusion}

The main result of the present paper is the proof that, considering the cosmological constant $\Lambda$ as an eigenvalue of a Sturm-Liouville problem allows one to obtain a set of admissible values for $\Lambda$. In the special case of a situation with discrete spectrum containing only one point, there is the possibility to strictly determine the cosmological conditions leading to this value of $\Lambda$ without recurring to the anthropic principle. We have shown in detail that this possibility is indeed realized in cyclic cosmological models when the potential of the matter field is similar to the potential of the scalar field. The dynamics of the universe in this case contains a sudden future singularity. As a byproduct we have also demonstrated that there is a remarkable connection between the Steinhardt-Turok cyclic models and sudden future singularity ones.
\bigskip

\noindent
{\bf Acknowledgments.}
The work by AVY has been supported by the ESF, project 4868 ``The cosmological constant as eigenvalue of Sturm-Liouville problem'', and the work by AVA has been supported by the ESF, project 4760 ``Dark energy landscape and vacuum polarization account'', both within the European Network ``New Trends and Applications of the Casimir Effect''. EE's research has been partly supported by MICINN (Spain), contract PR2011-0128 and projects FIS2006-02842 and FIS2010-15640, by the CPAN Consolider Ingenio Project, and by AGAUR (Generalitat de Ca\-ta\-lu\-nya), contract 2009SGR-994. EE's research was partly carried out while on leave at the Department of Physics and Astronomy, Dartmouth College, 6127 Wilder Laboratory, Hanover, NH 03755, USA.

\end{document}